\documentclass{article}

\newcommand{\bfz}{\textbf{z}}
\newcommand{\bfx}{\textbf{x}}

\usepackage[nonatbib]{neurips_2021}

\usepackage[utf8]{inputenc} 
\usepackage[T1]{fontenc}    
\usepackage{hyperref}       
\usepackage{url}            
\usepackage{booktabs}       
\usepackage{amsfonts}       
\usepackage{nicefrac}       
\usepackage{microtype}      
\usepackage{xcolor}         
\usepackage{graphicx}
\title{Analyzing High-Resolution Clouds and Convection using Multi-Channel VAEs}

\author{%
  Harshini Mangipudi\\
  University of California Irvine\\
  \texttt{harshinimangipudi@gmail.com} \\
  \And
  Griffin Mooers \\
  University of California Irvine \\
  \texttt{gmooers96@gmail.com} \\
  \And
  Mike Pritchard \\
  University of California Irvine \\
  \texttt{mspritch@uci.edu} \\
 \And
  Tom Beucler\\
  University of Lausanne \\
  \texttt{tom.beucler@unil.ch} \\
  \And
  Stephan Mandt \\
  University of California Irvine \\
  \texttt{mandt@uci.edu} \\
   
}

\begin{document}
\nolinenumbers
\maketitle

\begin{abstract}
Understanding the details of small-scale convection and storm formation is crucial to accurately represent the larger-scale planetary dynamics. Presently, atmospheric scientists run high-resolution, storm-resolving simulations to capture these kilometer-scale weather details. However, because they contain abundant information, these simulations can be overwhelming to analyze using conventional approaches. This paper takes a data-driven approach and jointly embeds spatial arrays of vertical wind velocities, temperatures, and water vapor information as three "channels" of a VAE architecture. Our "multi-channel VAE" results in more interpretable and robust latent structures than earlier work analyzing vertical velocities in isolation. Analyzing and clustering the VAE's latent space identifies weather patterns and their geographical manifestations in a fully unsupervised fashion. Our approach shows that VAEs can play essential roles in analyzing high-dimensional simulation data and extracting critical weather and climate characteristics. 

\end{abstract}

\section{Introduction}
\label{sect:intro}  
Long-term climate modeling is increasingly accurate, but any climate simulation high resolution enough to explicitly resolve kilometer-scale processes controlling clouds and storms not only requires high computational expense but also produces massive amounts of data. 
This large data volume overwhelms physically-informed methods traditionally used to better understand storm organization and how small-scale clouds connect to the large-scale circulation; even established dimensionality reduction techniques such as Principle Component Analysis (PCA)~\cite{wilks_2006} may fail to capture these nonlinear relations. In contrast, unsupervised learning, more specifically deep generative modeling ~\cite{goodfellow2014generative, crommelin2020resampling} can summarize explicitly-resolved storms without omitting fine-scale, nonlinear information.

Variational Autoencoders (VAEs) in particular are well-suited for dynamics discovery and understanding because of ability to learn meaningful latent representations of the data~\cite{mescheder2018adversarial, huang2018introvae, wu2017quantitative, pol2020anomaly}. VAE applications on climate and weather data have thus far been rare. Physical processes like a 2D laminar flow have been successfully represented by a VAE~\cite{eismann2017shape}. Some analysis has been conducted with VAEs on spatio-temporal earth dynamics and with state variables like temperature and potential vorticity data to cluster phenomena like the Polar Vortex~\cite{vae_earth_data, Polar_Vortex}. However, the training/test data applied to the VAEs in all these cases was highly idealized and smaller in dimension than a storm-resolving simulation output.

A \textit{single-channel} VAE architecture (SVAE), modeling velocity fields in high-resolution climate simulation has been implemented~\cite{mooers2020generative}; however, relationships between other variables known to be important controls for convection were ignored. To bypass the limit of using a single physical state variable input in the context of deep generative modeling for climate simulation analysis, we upgrade our network architecture to that of a three-channel VAE - temperature and water vapor in addition to vertical velocity. We hypothesize that this multi-variate approach can enable increasingly satisfying physical clustering through a more thermodynamically-informed latent space. 

\section{Methods}
\label{headings}

\subsection{Data and preprocessing}
\label{sim_info}

To provide our VAE with a multi-variate, high resolution simulation, we use output from the System for Atmospheric Modeling (SAM), a Storm-Resolving Model (SRM). We extract 1.5e6 30x128 snapshots or "images" of vertical velocity (m/s), temperature (K), and water vapor concentration (kg/kg) fields. The dimensionality of our images comes from the simulation design of an atmosphere with thirty vertical levels and 128 SRM columns embedded in each grid cell of a planetary-scale General Circulation Model (GCM) to explicitly resolve deep convection in thousands of embedded sub-domains~\cite{Grabowski1999}. The high-resolution convection data comes from all tropical areas ranging from latitudes of 20 degrees south to 20 degrees North during boreal winter. This geographically diverse output thus includes samples ranging from moist tropical rain forests to arid deserts, spanning both marine and continental zones. Each of the three variables is re-scaled separately from zero to one using standard normalization techniques.


\subsection{Model architecture and upgrades}
\label{vae_info}

We expand on a baseline VAE architecture designed in~\cite{mooers2020generative}. Our VAE is fully convolutional and has four layers in the encoder and decoder. To address the inherent rate-distortion trade-off native to VAEs~\cite{Alemi2018FixingAB} and align the latent variables of the model more closely with the output, we implement a linear annealing~\cite{Bowman2016GeneratingSF} in the Kullback–Leibler (KL) Divergence term on our VAE loss function (Equation~\ref{eq:elbo}). We initially weight the KL Divergence as 0 in order to prioritize learning in the encoder for a superior latent space but increase $\beta$ to 1 over 1600 epochs.

\begin{equation}
    \mathrm{ELBO}\left(\bfx; \theta, \phi, \bfz\right) = \mathbb{E}_{q_\phi(\bfz | \bfx)}\left[\log p_\theta(\bfx|\bfz)\right] - \beta \times D_{\mathrm{KL}}\left(q_\phi(\bfz | \bfx) \,||\, p(\bfz)\right), 
    \label{eq:elbo}
\end{equation}

In contrast to earlier work, we interpret temperature, water vapor, and raw vertical velocity fields as three “channels” of an image, and adopt a VAE architecture to jointly embed them. This also increases the compression from the input vector to a latent space of 1024 from roughly 4x in~\cite{mooers2020generative} to 10x. We also continue to use a SVAE trained on vertical velocity fields as a baseline for comparisons. 



\section{Results}
\label{results}

\paragraph{Latent space credibility and exploration}

\begin{figure}
\begin{center}
\begin{tabular}{c}
\includegraphics[width=\textwidth]{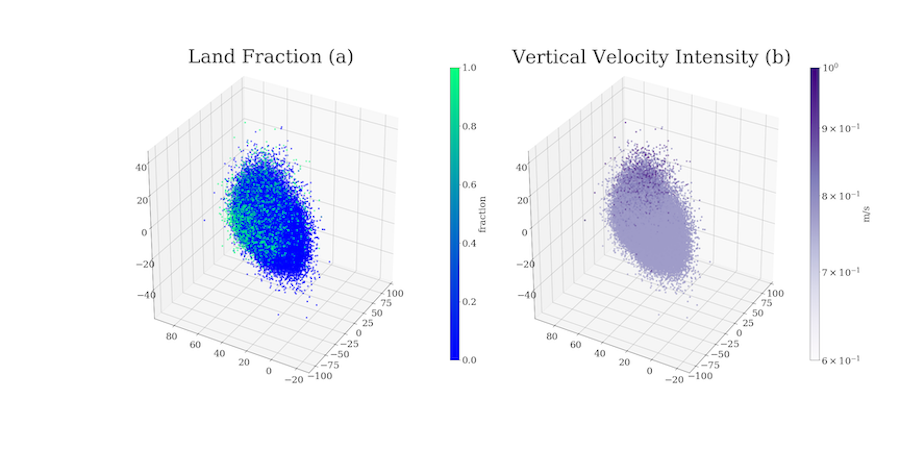}
\end{tabular}
\end{center}
\caption 
{ \label{fig:3D_Latent_Space}
The 3D projection of Principle Component Analysis of our multi-channel VAE latent space. We colorize by land/sea fraction and intensity of the vertical velocity field at each sample in the test dataset. The organization of this latent structure appears physically interpretable.} 
\end{figure}

By expanding the input to three high-resolution variables, we are asking substantially more of our encoder in terms of dimensionality reduction and feature extraction. There is no guarantee this architecture can create the same physically interpretable latent space as the SVAE in~\cite{mooers2020generative}. However, we can gain better understanding of the latent space by "colorizing" it by the magnitude of the corresponding vertical velocity field in the test dataset and its land fraction. In the tropics there are regions of intense convection, such as over the rain-forests and hot tropical waters of the Pacific warm pool; however, there will also be dry regions like deserts where deep convection will be suppressed. We want such known disparate convection types to be separated in some capacity on the latent space for it to be useful for physical analysis. When our latent space is colorized (Fig~\ref{fig:3D_Latent_Space}, a and b), we do see a clear sorting by magnitude, and convection over land preferentially appearing on different parts of the latent space than convection over the ocean. The full animated visual can be found at \href{https://drive.google.com/file/d/1plDwjMxbpCQ_ZwEqPbZLFvmtQSXZBT28/view?usp=sharing}{this link}. This suggests that the added complexity is not overwhelming the VAE, and we now ask: \textit{What are the advantages of the MVAE compared to the SVAE when it comes to anomaly detection and regime classification?}

\begin{figure}
\begin{center}
\begin{tabular}{c}
\includegraphics[width=\textwidth]{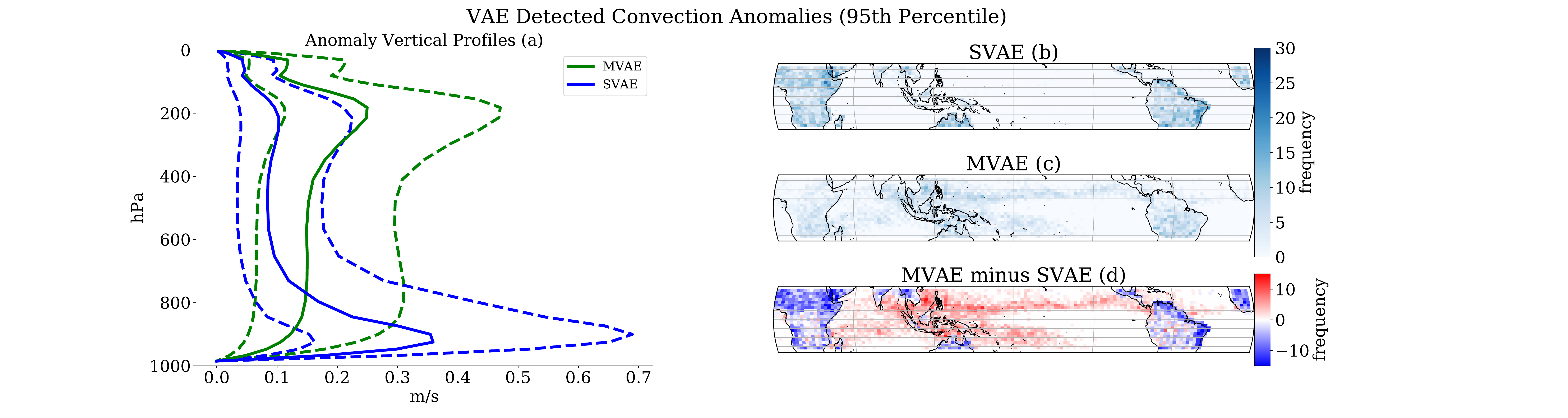}
\end{tabular}
\end{center}
\caption 
{ \label{fig:ELBO}
The vertical profiles (a, solid lines), 25th and 75th percentiles (a, dashed lines) and geographic distributions (b,c,d) of convection anomalies defined by our VAEs. We threshold the test data by the Evidence Lower Bound (ELBO) anomalies and examine only the top 5 percent of ELBO scores. The multi-channel VAE finds anomalies of different structure and geographic location than the single-channel VAE.} 
\end{figure}

\paragraph{Anomaly detection}
We determine the similarity between the trained VAEs and their respective latent spaces by first looking at density estimation properties. To test whether the MVAE adds value compared to the SVAE, we calculate the Evidence Lower Bound (ELBO) for both VAEs at each of the 1.5e6 samples in the test dataset. The higher this ELBO score the more anomalous the VAE finds the convection sample. We look at just the top five percent of ELBO anomalies for both the SVAE and MVAE. We then plot the frequency at which these anomalies occurred at each grid-cell in the tropics (Fig~\ref{fig:ELBO} b and c) and ask whether the results agree with domain knowledge of where geographically rare convective events tend to occur. Immediately we see different patterns lit up by the different VAEs. The SVAE hones in on continental zones, especially over more arid regions (Fig~\ref{fig:ELBO}b). The MVAE highlights areas over both land and sea (Fig~\ref{fig:ELBO}c), which is physically satisfying since rare storms are known to occur in both regions. Specifically, the MVAE finds deep convection over the tropical rain forests and the Inter-Tropical Convergence Zone to be most anomalous. But more significantly than just disagreeing about the geographic location of the most extreme anomalies, the two VAEs each find anomalies of very different vertical structures. In the median vertical profile in each anomaly group, the SVAE emphasizes strong convection near the surface of the earth, which is not typically recognized as an interesting phenomenon in the atmospheric sciences, whereas the MVAE focuses on convection deep in the upper Troposphere, a reassuringly familiar extreme tropical storm structure (Fig~\ref{fig:ELBO}a). 

\begin{figure}
\begin{center}
\begin{tabular}{c}
\includegraphics[width=\textwidth]{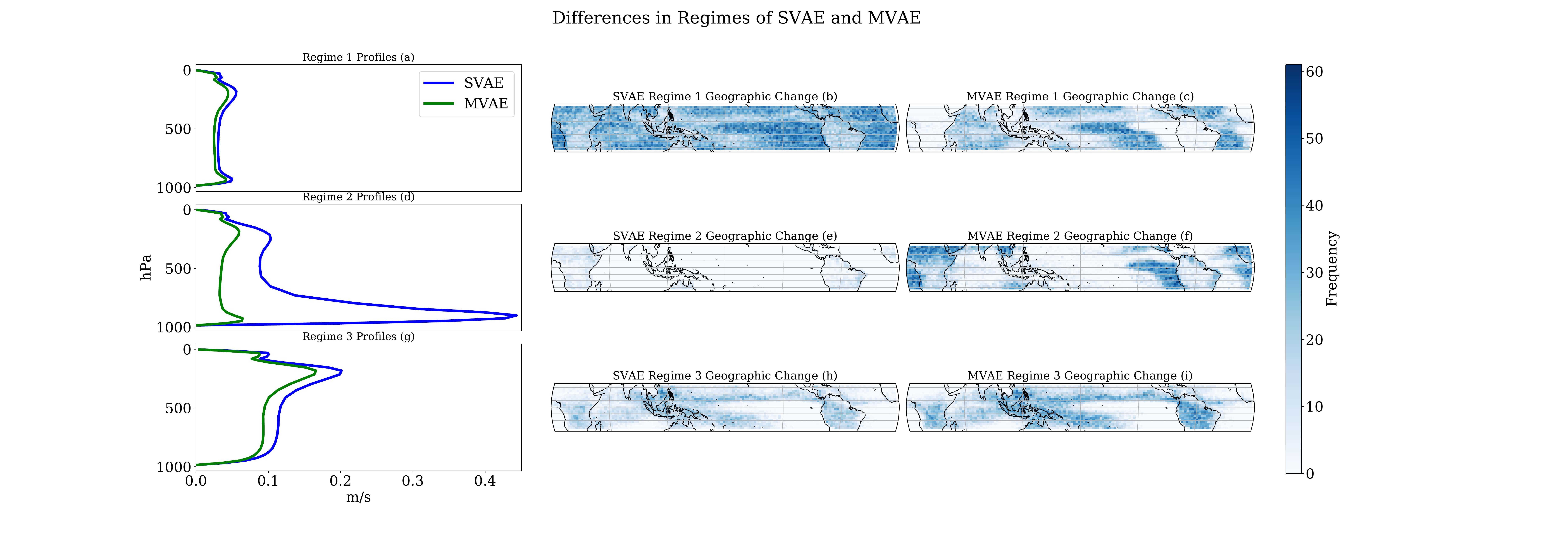}
\end{tabular}
\end{center}
\caption 
{ \label{fig:MultiVAE}
The results of K-Means clustering on the latent space of both VAEs separately. We find that three is the largest number of clusters in which each convective regime has both a distinct vertical profile (a,d,g) and distinct geographic pattern of frequency of occurrence (b,c,e,f,h,i). However, the three regimes found in each latent do not match suggesting unique organization in each VAE.}
\end{figure}

\paragraph{Data driven convection regimes}
To analyze the latent space in an objective fashion, we bring in another unsupervised learning algorithm: K-Means Clustering. We cluster separately on both latent spaces to determine if the latent structures are different enough for the MVAE to highlight different types of convective organization. Our analysis turns up three distinct clusters in both the latent spaces (Figs~\ref{fig:MultiVAE}). However, there are significant differences in the vertical profiles and geographic location of the regimes. The SVAE latent space successfully separates the deepest convection from shallower forms that occur in the subtropical ocean basins (Fig~\ref{fig:MultiVAE}a and g, blue lines). But the third cluster in the SVAE group, similar to the anomalies it finds (Fig~\ref{fig:ELBO}b), reflects the same puzzling mode of shallow convection over drier land areas (Fig~\ref{fig:MultiVAE}d, blue line). The MVAE clusters deep convection similarly to the SVAE but attractively does not make a distinct regime for the mode of dry, continental shallow convection (Fig~\ref{fig:MultiVAE} d,g green vs. blue). Instead, unlike the SVAE, the subtropical shallow convective mode is satisfyingly separated into two distinct sub-regions -- one in the Central Pacific/Atlantic, and another in the eastern parts of these ocean basins where especially distinct stratocumulus clouds are known to occur ~\cite{TrimodalCharacteristicsofTropicalConvection, tri_conv_Khouider, MAPES20063}. Since these two forms of low cloud have different characteristic horizontal eddy scales, we speculate that the inclusion of water vapor to the input of the MVAE down-weights the importance of the dry, shallow mode (Fig~\ref{fig:MultiVAE}d; green vs. blue) and allows a more satisfying physical clustering of convective regimes.


\section{Conclusion and broader impacts}
\label{conclude}
We explore the ability of a Variational Autoencoder (VAE) to find a physically interpretable representation of a multi-variate, high-resolution climate simulation. We find that adding multiple variables as "channels" to our VAE highlights drastically different convection anomaly types and geographic locations compared to a VAE trained only on a single variable. The different regimes identified by clustering both latent spaces suggest there is value to a multi-variate approach using both dynamic and thermodynamic variables to cluster convection regimes successfully in a fully unsupervised fashion. It is possible that this multi-variate approach would also be valuable for global storm-resolving climate simulations that produce even more data and augment traditional physically-informed approaches to clustering the cloud regime-dependence of climate change projections.


\section{Acknowledgements}
\label{thanks}
The authors thank the MAPS program and NSF grant 1633631, OAC-1835863, AGS-1734164, IIS-2047418, IIS-2003237 and IIS-2007719 for funding support and co-funding by the Enabling Aerosol-cloud interactions at GLobal convection-permitting scalES (EAGLES) project (74358), of the U.S. DOE Office of Biological and Environmental Research, Earth System Model Development program area. The link for training and analysis of the MVAE can be found at \href{https://github.com/HarshiniMangipudi/NeurIPS2021.git}{this link.}
Computational resources were provided by the Extreme Science and Engineering Discovery Environment supported by NSF Grant number ACI-1548562 (charge number TG-ATM190002).

\newpage
\bibliographystyle{abbrv}
\bibliography{main.bbl}

\end{document}